\documentclass[11pt]{article}
\usepackage{blois,epsfig}

\def\Journal#1#2#3#4{{#1} {\bf #2}, #3 (#4)}

\def\ApJS{\em Astrophys. J. Suppl. Ser.}
\def\AA{\em Astron. \& Astrophys.}

\def\be{\begin{equation}}
\def\ee{\end{equation}}
\def\bea{\begin{eqnarray}}
\def\eea{\end{eqnarray}}

\begin{document}
\vspace*{4cm}
\title{KLUN+ peculiar velocity survey}

\author{ M.O. Hanski }

\address{Laboratoire de Physique et Chimie de l'Environnement,\\
CNRS Orl\'{e}ans}

\maketitle\abstracts{
The enhanced Kinematics of the Local Universe (KLUN+) 
galaxy sample is a collection of galaxies suitable for Tully-Fisher (TF)
or Faber-Jackson (FJ) distance estimation. Here we extract a subsample
of 6229 KLUN+ galaxies closer than $80h^{-1}$~Mpc, and calculate their
distances and peculiar velocities with the Iterative Normalized Distance 
method. Within this method we can derive an analytical formula,
independent from the density inhomogeneities, for correcting 
the selection biases. The radial peculiar velocities can then be derived
from the redshifts and the corrected distances.
The  velocities are smoothed,
and the smoothed velocity field is used as a correction term at the next
derivation of normalized distances. This iterative procedure is repeated
until converging values are reached.
Here we present the resulting map of the radial peculiar velocity field at the 
$<80 h^{-1}$~Mpc environment. The infall patterns towards the main
galaxy clusters are clearly visible.
The color version of the map, other figures,
and animations are provided on the project web site.}

\section{Background}

Peculiar velocities give direct observational information of 
the gravitationally induced structure formation in action. 
A peculiar velocity catalog can be used for studying cosmological
parameters, bulk flows, and the theory of structure formation.
The data can also be used for setting the initial conditions for
constrained simulations. 

The largest published peculiar velocity catalog has been the
MARK III (Willick et al.\ \cite{wi97}), which is a compilation
of six TF surveys and a sample of ellipticals with $D_n$-$\sigma$
distances. There the spiral samples are converted to a common system
and combined with the ellipticals. In the KLUN+ survey we use the
same strategy. Our attempt is to use all the available data
suitable for TF or FJ studies;
the sample is constantly updated by the new published data, most
notably those provided by our observational program at the Nan\c{c}ay 
radiotelescope (Theureau \cite{th03}). Data obtained from different sources are carefully
homogenized to a common system. Presently the sample contains about
20\,000 galaxies, with magnitudes in B, I, J, H, and/or K bands,
radial velocities, and rotational velocities (spirals) or central
velocity dispersions (ellipticals).
For studying the peculiar velocity field, we
extract a magnitude limited sample, and exclude galaxies having
large errors in the observables. In the present study, the
$80 h^{-1}$~Mpc sphere around us, we finally have 6229 KLUN+ galaxies. 
In MARK III the number of galaxies within this distance is only 3442.

\section{Iterative Normalized Distance method}

The derivation of peculiar velocities requires that we
obtain galaxy distances using redshift independent distance indicators, such as the
TF or the FJ relation.
These have to be corrected for the selection biases, the errors
in the derived distances caused by the scatter of the relations 
and the detection limits.
In KLUN studies, a ``normalized distance'' method has been used for
correcting the bias (Theureau et al. \cite{th97}). There the distances are scaled so that 
for all objects at a fixed normalized distance the amount of bias is the same.
An analytical correction formula can then be applied (Theureau et al. \cite{th98}).

For obtaining the peculiar velocities, an ``Iterative Normalized Distance''
method can be
used. There the  main steps are (only the TF relation is mentioned, but
the same procedure works for the FJ relation):
\begin{enumerate}
\item Calculate the absolute magnitudes and the normalized distances 
using the kinematical (redshift) distances.
\item Calculate the TF relation
using the unbiased part of the normalized distance diagram.
\item Use the unbiased TF relations and the analytical bias correction
formula for estimating real space galaxy distances beyond the 
unbiased plateau limit.
\item Interpolate the peculiar velocity field in a Cartesian grid in the redshift
space by smoothing the individual peculiar velocities given by the
bias corrected TF distances. 
\item Go back to step 1, and use the corrected kinematical distances by
subtracting the smoothed peculiar velocity field values from the redshift 
velocities.  
\end{enumerate}
This loop is repeated until converging values for the peculiar velocities
are obtained. 5--10 iterations are sufficient. 

We then use a bootstrap technique to exclude some of the objects
causing large uncertainties. From 50 bootstrap replications of the original
sample, we seek for any correspondence of a presence/absence of a galaxy
in a sample, and large deviations in the peculiar velocity field at the
corresponding region. We set a threshold for acceptable deviation,
and exclude about 2\% of the original sample, causing uncertainties
larger than this limit. Finally, the peculiar velocity field is
recalculated with the reduced sample.

\begin{figure}
\psfig{figure=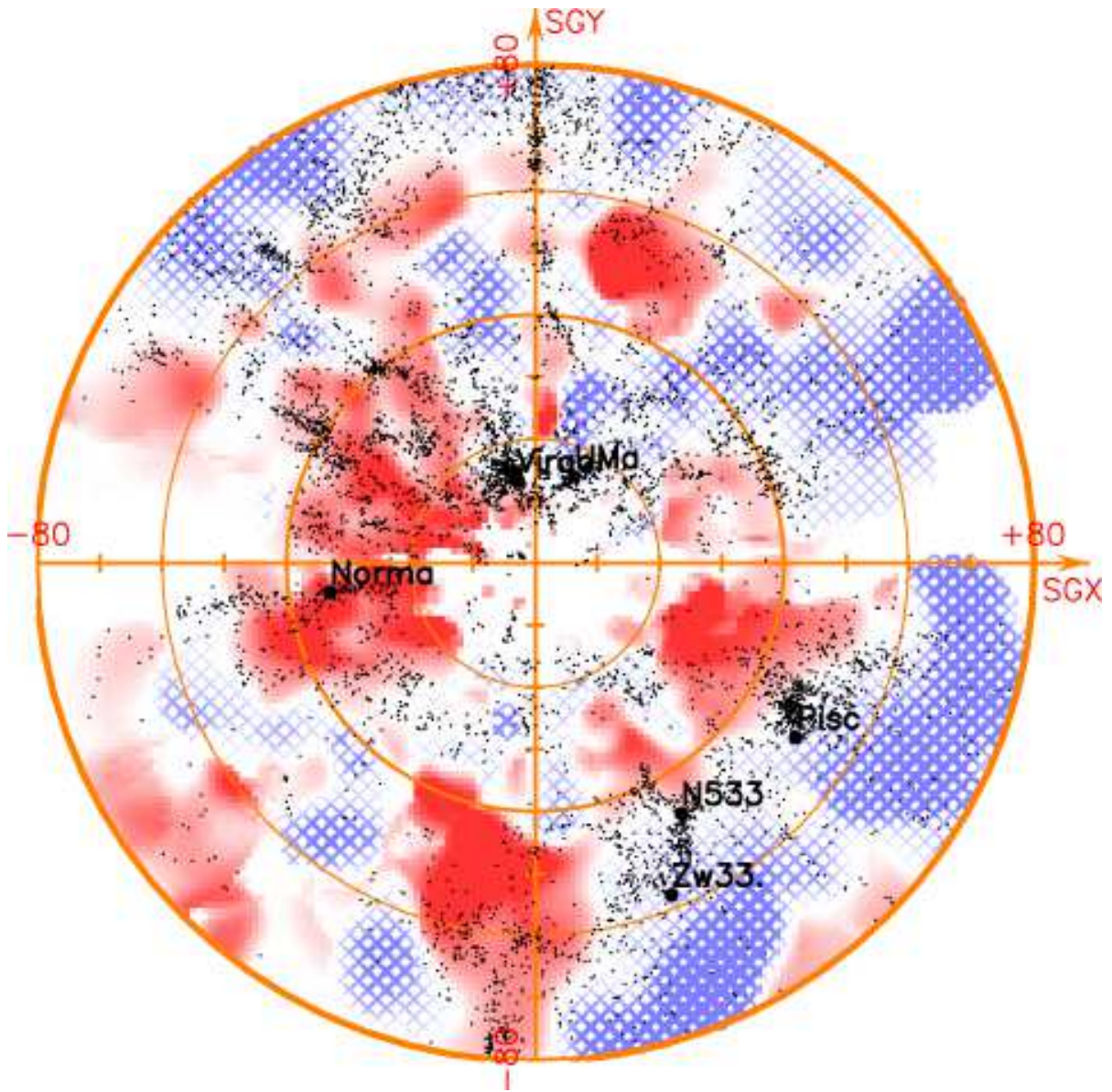,width=6.25in}
\caption{Radial peculiar velocity field in the supergalactic plane.
The coordinates are 
in ``real space'', i.e.\ redshift distances corrected for the smoothed peculiar
velocity field, in units of $h^{-1}$~Mpc.}
\end{figure}

\section{Results}

Figure 1 shows a map of the smoothed radial peculiar velocity field
in a disc centered on the supergalactic plane. Our galaxy is in the center.
The thickness of the disc is zero in the center and increases towards the edge.
The opening angle is $15^\circ$ so that at the edge the thickness
is roughly 20~$h^{-1}$~Mpc. The hatched regions correspond to negative
velocities, these regions are falling towards us. The solid regions are
moving away. The darkness of the colors corresponds to the amplitude
of the velocity. Regions where the velocity is not defined are white.
The black dots are all the galaxies in the HYPERLEDA\footnote{http://leda.univ-lyon1.fr/} database with measured
redshifts (there are about 32\,000 galaxies in HYPERLEDA with redshift distances $<$~80~$h^{-1}$~Mpc).  Circles mark some well
known clusters --
Virgo, Ursa Major, Norma, Pisces, N533, and Zwicky33.
The map is in real space coordinates. 

It is worth noting that in the map one observes both the front and backside infall patterns around the main
superclusters and structures. It is particularly obvious  for the regions of Virgo, 
Perseus-Pisces, N533, Norma, or even along the Great Wall, which spans from
10 o'clock, 60~$h^{-1}$~Mpc to 12 o'clock, 80~$h^{-1}$~Mpc.

In order to present the peculiar velocity field of the whole
80~$h^{-1}$~Mpc sphere we have constructed animations showing the
disc rotating around one of its axes. These animations, and more
figures are available at the website http://klun.obs-nancay.fr/VPEC/vpec.html.

The results presented here are
preliminary; the details of the method and the final results will be published shortly (Hanski et al.\
\cite{ha03}).

\section*{Acknowledgments}
This work has been supported by the Academy of Finland, project
``Mapping and numerical modeling of the local
galaxy universe'', and the R\'{e}gion Centre.

\end{document}